# Application-oriented strain-hardening engineering of high-manganese steels

Christian Haase[1], Franz Roters[2], Angela Quadfasel[3]

[1]*Steel Institute IEHK, RWTH Aachen University, Intzestr. 1, Aachen 52072, Germany,*
*christian.haase@iehk.rwth-aachen.de*
[2]*Max-Planck-Institut für Eisenforschung GmbH, Max-Planck-Str. 1, Düsseldorf 40237, Germany*
[3]*Institute of Metal Forming, RWTH Aachen University, Intzestr. 10, 52072 Aachen, Germany*

**Abstract**

The outstanding mechanical properties of high-manganese steels (HMnS) are a result of their high strain-hardenability. That is facilitated by strong suppression of dynamic recovery, predominant planar glide, and the activation of additional deformation mechanisms, such as transformation-induced plasticity (TRIP) and twinning-induced plasticity (TWIP). However, depending on the final application, strongly differing requirements on the mechanical properties of HMnS are of relevance. In order to design HMnS for specific applications, multi-scale material simulation under consideration of the processing conditions that allows for prediction of the final mechanical properties is required, i.e. strain-hardening engineering based on integrated computational materials engineering (ICME). In this work, we present an approach that employs alloy selection by stacking-fault energy calculations, which enables activation/suppression of specific deformation mechanisms. Tailored manufacturing, e.g. by thermo-mechanical treatment, severe plastic deformation or additive manufacturing, provides possibilities to make use of the strain-hardenability during and after processing to define the mechanical properties. A computational approach for alloy and process design will be discussed.

**Keywords**

Strain hardening, mechanical properties, modelling, simulation

## 1 Introduction

High-manganese steels (HMnS) including transformation-induced plasticity (TRIP), twinning-induced plasticity (TWIP) and slipband refinement-induced plasticity (SRIP) steels received very high attention in both academia and industry. This is mainly motivated by their mechanical properties that are superior to any other class of steel with respect to the combination of tensile strength and elongation to fracture [Bouaziz 2011]. Nevertheless, wide industrial application has been impeded so far due to limited tailoring of the materials' properties according to the requirements of specific applications. For instance, application of fully recrystallized HMnS in crash-relevant automobile components was found to be unsuitable, since the high formability and strain-hardening capability can hardly be utilized [Bambach 2016].

During the last years, numerous alloy design strategies and processing methods have been explored to vary the mechanical properties of manganese-alloyed steels, e.g. [Haase 2014, Haase 2016, Kies 2018]. As shown in Figure 1, these properties can be adjusted within a wide range. However, in order to engineer the strain hardenability of HMnS for specific applications, robust computational methods that allow for precise prediction and efficient design of the mechanical properties are required. In the present study, we present a physics-based integrated computational materials engineering (IMCE) methodology that takes the chemical composition, processing as well as the operating conditions into account. The methodology was employed for the design, production and application of automobile crash-boxes made of an X30MnAl23-1 TWIP HMnS.

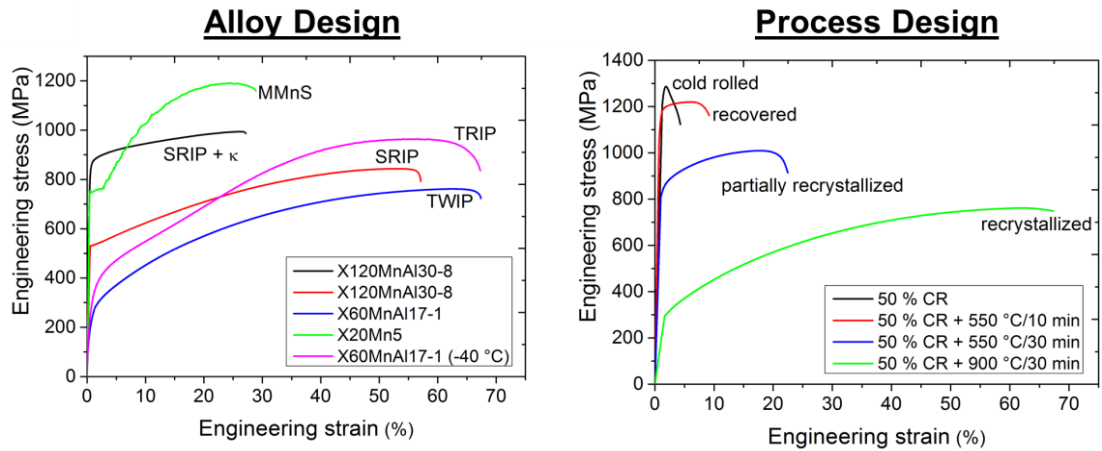

Figure 1: Mechanical properties of HMnS and medium-manganes steels (MMnS) tailored based on alloy design (left) and process design (right).

## 2 Approach and results

The integrative simulation approach used for application-oriented strain-hardening engineering is shown schematically in Figure 2. The initial stacking fault energy (SFE), which determines the active deformation mechanisms, was calculated based on a combined ab initio and CALPHAD approach [Saeed-Akbari 2009]. The SFE served as input for a physics-based crystal plasticity-finite element model (CP-FEM) that allows for consideration of dislocation slip as well as deformation twinning and martensitic phase transformation [Steinmetz 2013, Wong 2016]. As a basis for macroscopic simulation of the deformation behaviour under operation conditions (e.g. crash simulation using FEM), the material development during processing was predicted using the hardening model (deformation, e.g. rolling) and a softening model via cellular automaton (CA) approach (recrystallization) [Haase 2015]. Finally, the output of the crash simulation using FEM, such as buckling behaviour, force-distance curves and the resulting deceleration, enables critical evaluation of the simulation approach and optimization of the alloy and process concepts.

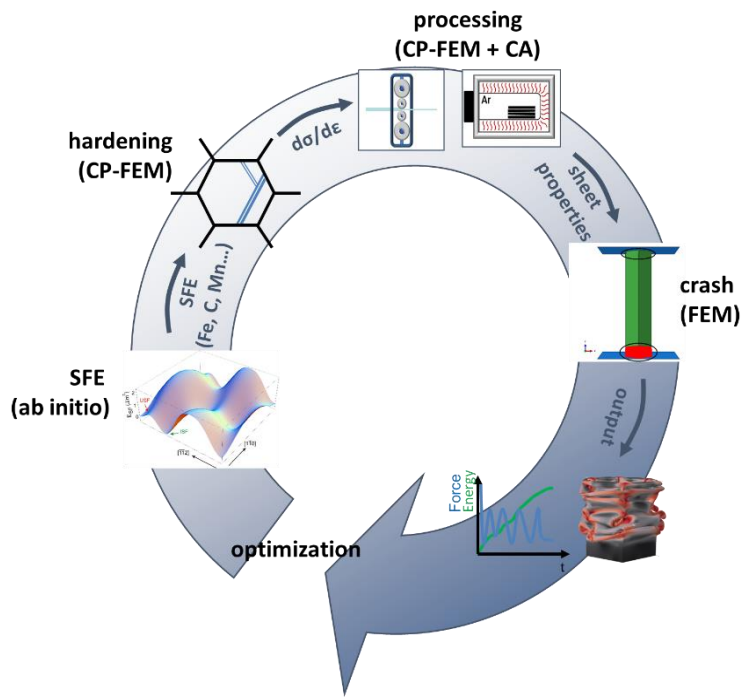

Figure 2: Simulation approach for application-oriented strain- hardening engineering.

Figure 3 presents the results of the simulation approach used. The chemical composition resulted in a calculated SFE of 25 mJ/m$^2$, which indicates plastic deformation by dislocation slip and deformation twinning, whereas the TRIP effect is suppressed. The processing conditions, i.e. combination of strain hardening due to cold rolling (CR) and softening due to recovery annealing (RV), had a significant influence on the sheet properties of the materials (cf. Figure 3, middle). On the one hand, the yield stress was drastically enhanced with increasing rolling degree, whereas the recovery processes during annealing facilitated regained ductility [Haase 2014]. On the other hand, high initial deformation degrees by cold rolling introduce high dislocation and twin densities and thus, reduce the strain-hardening capability during further deformation. As a consequence of the high initial flow stress after cold rolling with a thickness reduction of 50%, the crash distance required to absorb an impact energy of 5 kJ was strongly reduced, as compared to the fully recrystallized (RX) and 30%/40% CR+RV material (Figure 3, bottom). However, it must also be noted that the high force amplitudes in the force-distance curves of the recovery-annealed states lead to pronounced deceleration and thus, to increased probability for whiplash injury of passengers. To overcome this drawback the geometry of the cash-box can be adjusted. In this case, the increasing of material strength can be used to produce light weight crash-boxes. In comparison with experimental data, both the simulated sheet properties of the differently processed material as well as the simulated deformation behaviour during crash testing showed excellent agreement. Therefore, the developed simulation approach allows for accurate prediction of the material development during processing and of the behaviour under operation conditions.

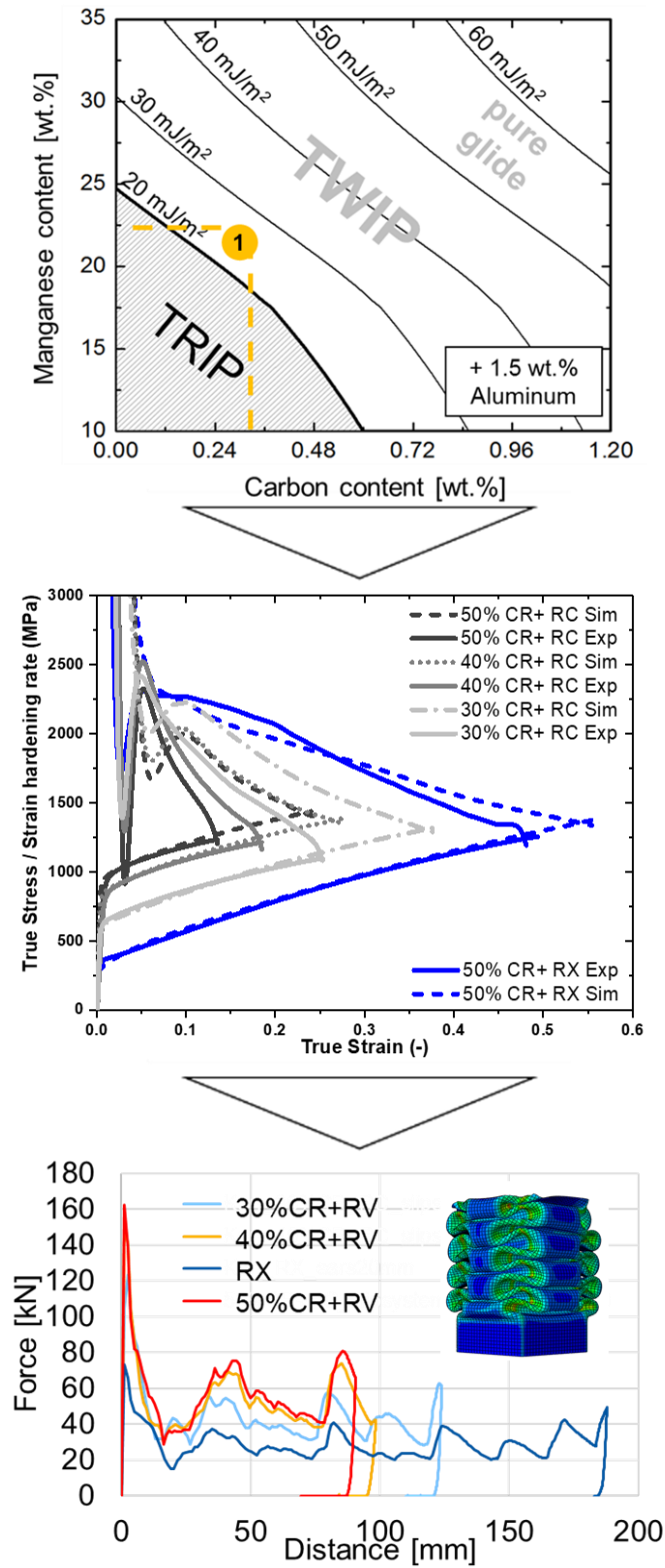

Figure 3: Results of the SFE calculation of the investigated X30MnAl23-1 steel (top), the corresponding simulated tensile properties depending on the processing route (middle) and the predicted behaviour during drop-tower testing (bottom).

## 3  Summary

In the present study, an ICME methodology was developed and employed in order to predict the behaviour of an X30MnAl23-1 TWIP steel during processing and application in energy-absorbing crash-boxes. Precise calculation of the SFE and identification of the activated deformation mechanisms allowed for accurate simulation of the mechanical properties of HMnS under consideration of processing conditions during thermo-mechanical treatment. Further coupling with a FEM model enabled the description of the macroscopic deformation of the crash-boxes. Due to the excellent agreement between simulations and experiments, the simulation approach serves as a basis for efficient optimization of alloy and process design strategies without the necessity of a large number of experimental trial and error approaches.

**Acknowledgement**

This work has been funded by the German Research Foundation (DFG) within the collaborative research center (SFB) 761 "Steel - ab initio. Quantum mechanics guided design of new Fe based materials". The authors would like to thank all researchers contributing to Cloud I "Strain-hardening engineering" within in the SFB 761.